\documentclass[12pt,a4paper]{article}
\usepackage{amsfonts,amsmath,amsthm,amssymb,mathtext,latexsym}
\usepackage{graphicx}
\usepackage[T2A]{fontenc}
\usepackage[utf8]{inputenc}
\usepackage[english]{babel}

\usepackage{csquotes}

\newtheorem{thm}{Theorem}[section]

\theoremstyle{definition}

\theoremstyle{remark}

\numberwithin{equation}{section}

\begin{document}
\title{To the Issue of Justification of the Absolutely Continuous Spectrum Eigenfunctions Asymptotics
in the Problem of Three One-Dimensional Short-Range Quantum Particles with Repulsion.}
\author{A.M. Budylin, S.B. Levin}
\date{}
\maketitle
\tableofcontents

\section{Introduction}
In the work~\cite{BL1} (based on \cite{BMS}-\cite{BMS2})
there were introduced for the first time
asymptotic formulas at infinity in coordinates for absolutely continuous
spectrum eigenfunctions uniform at angular variables in the case of a
quantum system of three one-dimensional particles with short-range repulsive
pair potentials. The mentioned asymptotic formulas were obtained
non-strictly within the framework of a fairly subtle heuristic analysis.
Nevertheless the significance of such formulas in the context of the
absence of proofs of the existence and uniqueness solution theorem
for the given scattering problem was so essential that in the
work~\cite{BLNO} quite a successful attempt of numerical modelling was made,
based on the results of the work ~\cite{BL1}.

Continuous spectrum eigenfunctions contain asymptotic contributions
of different kinds of arising. As was shown (though on the heuristic level)
in~\cite{BL1}, in the case of one-dimensional particles the contributions
of different kinds of arising have in the sense of weak asymptotics the same type of decrease
at infinity in configuration space. The difference is only in
there being or not singularities at angular variable in the diverging
circular wave coefficient. Therefore a strict mathematical analysis of the
situation seems especially complicated.

Exactly, as was shown, for example, in~\cite{BL1}, the wave generated by
three-particles process (scattering on the center) decreases as $x^{-1/2}$ with a
smooth amplitude. Contributions generated by two-particles scattering
processes (multiple re-scattering included) keep decreasing as $x^{-1/2}$,
but already with a singular at angular variable (in the sense of weak
asymptotics) amplitude.

Our goal is to describe these contributions on a strict mathematical level.
With the aim in mind we give priority to studying resolvent limited values
on absolutely continuous spectrum.

Considering the subject of this problem we must mention the works~\cite{Mur}
and~\cite{PSS}, where the absence of the
Schroedinger operator singular continuous spectrum for a broad class of
pair-wise interactions potentials is asserted. It means, in particular,
that the limit values of the resolvent
\begin{equation*}
  w\text{-}\lim\limits_{\varepsilon\searrow0}R(E\pm
i\varepsilon)\,,\quad E\in(0,\infty)
\end{equation*}
of the operator considered here exist as absolutely continuous functions
of variable $E$.

However, the Moore technique does not allow to find these limit values
and moreover, to distinguish from them a full system of absolutely
continuous spectrum eigenfunctions.

Within the frameworks of our approach which in principle is close to the
fundamental work ~\cite{Fad}, the above mentioned limit values are
calculated explicitly.

It is necessary to stress that the limitation of considerations for the case
of finite pair potentials does not lead to a problem simplification in
principle, as the interaction potential of all three particles
remains non-decreasing at infinity but allows to set aside a certain
number of technical details.

\section{Preliminaries}
\subsection{ Jacobi Coordinates. Reduction.}
In the initial setting we consider a non-relativistic Hamiltonian $H$
\begin{equation}
\label{ISHOD-H}
H\psi=-\Delta \psi +\frac{1}{2}\sum_{1\leqslant i\ne j\leqslant 3}v(z_{i}-z_{j})\psi\,,
\quad z_{i}\in \mathbb{R}\,,
\quad \boldsymbol{z}=(z_{1},z_{2},z_{3})\in \mathbb{R}^{3}\,,\quad
\psi=\psi(\boldsymbol{z})\in\mathbb{C}
\,,
\end{equation}
$\Delta$ is a Laplace operator in $\mathbb{R}^{3}$, $v$ is an even\footnote{
  the requirement of evenness is not essential but is dictated by physical
  setting of the problem.} function 
$\mathbb{R}\to
[0,+\infty)$, defining two-particles interaction\footnote{in the setting a
  system of three quantum equal mass
particles is considered.}.

For a better illustration we will regard the function $v$ as a finite
(and integrable) one. In the case the essential self-adjointness of the
operator $H$ in the square integrable functions space is well known, see,
for example, ~\cite{Fad}. We will describe more precise requirements to $v$ a
bit later.

The extension of the results obtained here to the case of arbitrary
fast decreasing at infinity pair potentials, as well as Coulomb
potentials, is to be discussed in subsequent publications.

\begin{figure}[htbp]
\label{krest}
\begin{center}
     \includegraphics[width=6cm]{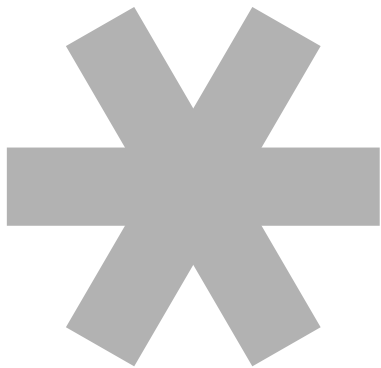}
\caption[]{The compact part of the operator $H$ potential support.}
\end{center}
\end{figure}

To separate the center of mass motion, we confine Hamiltonian on the
surface $\Pi$, defined by the equation $\sum z_{i}=0$. Allowing here a certain
negligence, we will hereafter define by $\Delta$ a Laplace-Beltrami operator on
the plane $\Pi\subset \mathbb{R}^{3}$. As this takes place, on $\Pi$ we can use any pair
$(x_{i},y_{i})\,,\; i=1,2,3,$ of the so called Jacobi coordinates, defined by
the equations
\begin{equation*}
x_{i}= \tfrac{1}{\sqrt{2}} (z_{k}-z_{j})\,, \quad y_{i}=\sqrt{\tfrac{3}{2}}z_{i}\,,
\end{equation*}
the indices  $(i,j,k)$ here form even permutations. In view of the
orthonormality of Jacobi coordinates and invariance of the
Laplace-Beltrami operator we have
\begin{equation*}
\Delta= \frac{\partial^{2}}{\partial x_{i}^{2}}+\frac{\partial^{2}}{\partial y_{i}^{2}}\,,
\end{equation*}
and our operator takes the form
\begin{equation}
\label{osnovnoj-H}
H=-\Delta +V\,,\qquad V=\sum_{i=1}^{3} v_{i}\,,\quad v_{i}=v(x_{i})\,.
\end{equation}
Note that the support of potential $\sum v_{i}$ lies in the infinite cross domain,
the compact part of which is shown on the picture~\ref{krest}.



Define the resolvent of the operator $H$
\begin{equation*}
R (\lambda)= (H-\lambda I)^{-1}\,,\quad \lambda\notin[0,+\infty)\,,
\end{equation*}
From now on $I$ is an identity operator. The resolvent $(H_{0}-\lambda I)^{-1}$ of
the unperturbed operator $H_{0}=-\Delta$ will be defined by $R_{0} (\lambda)$.

\subsection{Ultimate Absorption Principle}

We are interested in the scattering problem within the framework of a
so called stationary approach. In this case studying of wave operators is
replaced by studying of the limit values $R(E\pm i0)$ of the resolvent 
of the operator in question, when a spectral parameter $\lambda$ approaches
the real axis $\lambda\rightsquigarrow
E\pm i0\,,\; E\in (0,+\infty)$).
The proof of the existence of such limit values in a suitable topology
is the very essence of the ultimate absorption principle. Since the
existence of resolvent limit values is established,
studying of wave operators themselves follows the known standard scheme,
see, for example, ~\cite{Yaf},\cite{RS3}.

As we will see, absolutely continuous spectrum eigenfunctions $\psi^{\pm}(x,E)$ can
be reconstructed by the formula 
\begin{equation*}
  R(x,y|E\pm i0)\underset{y\to\infty}{\sim}\psi^{\pm}(x,E) \cdot \frac{e^{\pm
i\sqrt{E}|y|}}{\sqrt{|y|}}\,,
\end{equation*}
compare, for example,~\cite{MF}.

\subsection{Equipped Hilbert Space Method}
The point in question is exactly about one of the main methods of topology
introduction, in which limit values of the resolvent $R (E
\pm i0)\,,$ are
considered, see ~\cite{GelVil},\cite{RS3}.

In the case into the main Hilbert space $\mathcal{H}$  a certain Banach space $\mathcal{B}$
is continuously embedded, which in its turn allows to embed $\mathcal{H}$ into
the conjugate to $\mathcal{B}$ space $\mathcal{B}^{*}$ and next to prove that $R (E\pm i\varepsilon): \; \mathcal{B}\to \mathcal{B}^{*}$
has a continuous extension at $\varepsilon \downarrow 0$.

Thus generalized eigenfunctions in this case are treated as elements of
the space $\mathcal{B}^{*}$, while the main object of study becomes a scalar product
$(R (E\pm i\varepsilon)\varphi,\varphi)\,,\; \varphi\in \mathcal{B}\,,$ at  $\varepsilon \downarrow 0$.

Hereafter, to be definite, we will limit ourselves to the case $\mathrm{Im}\lambda \downarrow~0$.

\subsection{Friedrichs -- Faddeev Model}

In the Friedrichs - Faddeev model, see ~\cite{Fad},\cite{Yaf},
within the framework of the ultimate absorption principle an unperturbed
operator is treated as an operator of multiplication by an argument, while
the perturbation represents an integral operator with a smooth kernel.
For applications it is as a rule connected with a transition to a
momentum representation. Studying of resolvent analytical singularities
in a momentum representation can prove to be easier than studying of its
coordinate asymptotics. On the whole, we adopt this point of view too.

One of the main features of Faddeev approach~\cite{Fad} to many-body
scattering problems is a transition of operator resolvent to a so called
$t$- operator $T (\lambda)$, connected to the resolvent $R (\lambda)$ by the identity
\begin{equation*}
T (\lambda)=V-VR (\lambda)V\,,
\end{equation*}
where  $V$ is a perturbation of the operator $H_{0}$. To study the operator $T (\lambda)$
is considered the equation 
\begin{equation*}
T (\lambda)=V-VR_{0} (\lambda)T (\lambda)\,,
\end{equation*}
where the resolvent is reconstructed according to the equality
\begin{equation*}
R (\lambda)=R_{0} (\lambda)-R_{0} (\lambda)T (\lambda)R_{0} (\lambda)\,.
\end{equation*}
In Friedrichs - Faddeev model an unperturbed resolvent
``on continuous spectrum'' serves as an operator of multiplication by a
function with a pole. As a sequence, the kernel of the operator $T (\lambda)$ is
assumed to be H\"older continuous. The matter is that while describing
limit values of the above mentioned scalar product $(R (E\pm i0)\varphi,\varphi)$,
there naturally arises a singular Hilbert integral so H\"older functions,
contrary to functions only continuous, form function algebra, on which a
singular Hilbert integral is defined as a bounded operator.

It is exactly the circumstance that will define our choice as well of
the supplementary space, in the topology of which the resolvent values
limited in the weak sense will be considered. Such space in
an momentum representation will be the space
 $H^{\mu,\theta} (\mathbb{R}^{2})$ $(\frac{1}{2}<\theta<1,\frac{1}{2}<\mu<1)$
of H\"older functions with a norm
\begin{equation}
\label{geld-norma}
\|f \|_{\mu,\theta}=\sup\limits_{\xi,\eta} (1+|\xi|^{1+\theta})\Bigl( |f(\xi)|+
\frac{|f (\xi+\eta)-f (\xi)|}{
|\eta|^{\mu}}\Bigr)\,.
\end{equation}
The function space in coordinate representation, Fourier images of which
belong to  $H^{\mu,\theta} (\mathbb{R}^{2})$, we will define by $\hat{H}^{\mu,\theta}$.

In the work ~\cite{Fad} contributions accounting from particle
perturbations into the complete resolution was carried out on the basis
of the analysis of so called Faddeev equations.

The analysis of resolvent singularities in a momentum representation is
conveniently carried out within the framework of a so called
alternating Schwartz method, see ~\cite{Mor},\cite{BB1},\cite{BBsin}.
Note that the known Faddeev equations, see  ~\cite{Fad}, can be
interpreted as well as an alternating method version.

\subsection{Alternating Schwartz Method}
Naturally, first of all there stands out a possibility of separation of
variables for a partial Hamiltonian
\begin{equation}
\label{chastichnyj-H}
H_{i}=-\Delta+v_{i}\,.
\end{equation}
Thereby, the existence of limit values of the resolvent
$R_{i} (\lambda)=
(H_{i}-\lambda I)^{-1}$
is easily controlled, or rather see below section~\ref{asy-Gamma1}.

As a sequence, there rises a question of accounting of such contributions
into the resolvent $R (\lambda)$ of the complete Hamiltonian $H$ with a total
potential $\sum v_{i}$.

The scheme of such accounting is known in literature under the name of
an alternating Schwartz method, see~\cite{Mor},\cite{BBsin}.
Recall the main principles of the abstract version of the scheme, as
a number of its details plays an important part in understanding of
our computations.

Denote by $\{G_{i}\}|_{i=1}^{n}$ a certain number of linear operators in a complex vector
space $\mathcal X$. Define the operator
\begin{equation*}
G=\sum_{i=1}^n G_i\,.
\end{equation*}
Assuming that all operators $I-G_i$ are bijective, put
\begin{equation}
\label{gamma-i}
I-\Gamma_{i}=(I-G_i)^{-1}\,.
\end{equation}
The operators $\Gamma_{i}$ bear a name of reflection operators.

The principle of the alternating Schwartz method resolves itself
into the following
\begin{description}
\item[1)]
If the operator matrix
\begin{equation}
\label{L-big}
\boldsymbol{L}=\begin{pmatrix}I&\Gamma_1&\ldots&\Gamma_1\\
                             \Gamma_2&I&\ldots&\Gamma_2\\
                             \vdots&\vdots&\ddots&\vdots\\
                             \Gamma_n&\Gamma_n&\ldots&I
\end{pmatrix}
\end{equation}
is a bijection of the space ${\mathcal X}^n$, then the operator $I-G$ is a bijection
of the space $\mathcal X$.

\item[2)]
Denoting by  $\mathrm{diag} (\Gamma_{1},\ldots \Gamma_{n})$
a corresponding diagonal matrix. In conditions of bijectivity~\eqref{L-big}
we will define $\boldsymbol{\gamma}$ as a solution of the equation
\begin{equation}\label{psi-gamma}
\boldsymbol{L}\cdot\boldsymbol\gamma=\mathrm{diag}(\Gamma_1,\ldots\Gamma_n)
\end{equation}
and introduce linear operators  $\gamma_{ij}$ in $\mathcal X$ as components
of the matrix $\boldsymbol{\gamma}$. The latter can be also characterized by
the identities
\begin{equation}
\label{gammaij}
\gamma_{ij}=\Gamma_i\bigl(\delta_{ij}I-\sum_{k\ne i}\gamma_{kj}\bigr)=
\bigl(\delta_{ij}I-\sum_{k\ne j}\gamma_{ik}\bigr)\Gamma_j\,.
\end{equation}

Then the reflection operator $\Gamma$, defined by the equation 
\begin{equation}
\label{gamma-big}
I-\Gamma=(I-G)^{-1}\,,
\end{equation}
is given in terms of $\gamma_{ij}$ by the formula
\begin{equation}
\label{gamma-gamma}
\Gamma=\sum_{1\leqslant i,j\leqslant n}
\gamma_{ij}\;.
\end{equation}

\item[3)]
  Of special significance is the fact that the components $\gamma_{ij}$ can be
  reconstructed by the operator  $\Gamma$  by the formulas
\begin{equation}
\label{gamma-vosstanovlenie}
\gamma_{ij}=-\delta_{ij}G_j-G_i(I-\Gamma)G_j\,.
\end{equation}

\item[4)]
At $n=2$ the operator
$\boldsymbol{L}$ will be a bijection $\mathcal{X}^{2}$ then and only then, when the operator
$(I-\Gamma_1\Gamma_2)$ will be a bijection $\mathcal{X}$.

At that
\begin{equation}
\label{gamma-n2}
I-\Gamma=(I-\Gamma_2)(I-\Gamma_1\Gamma_2)^{-1}(I-\Gamma_1)\,.
\end{equation}
\end{description}
Detailed proof of all the facts can be found, for example, in ~\cite{BBsin}.

Note one more property which was not described in ~\cite{BBsin}. To be
exact, the assertion of the fourth section can be generalized also
for the case of an arbitrary number of perturbations:
\begin{description}
\item[5)] if the operator $I-G$ together with the operators 
  $I-G_{i}$ is a bijection of the space $\mathcal{X}$, then the operator $\boldsymbol{L}$ will be a
  bijection of the space $\mathcal{X}^{n}$ (with all ensuing consequences).
\end{description}
Indeed, the components of the inverse to $\boldsymbol{L}$ operator, to be denoted by $\omega_{ij}$,
must satisfy the equation
\begin{equation*}
\omega_{ij}+\Gamma_{i}\sum\limits_{k\ne i}\omega_{kj}=\delta_{ij}I\,,
\end{equation*}
which can be re-written in the form of
\begin{equation*}
\omega_{ij}= (I-G_{i})\delta_{ij}+G_{i}\sum_{k}\omega_{kj}\,.
\end{equation*}
But the sum  $\omega_{j}=\sum\limits_{k}\omega_{kj}$ 
is uniquely defined by the equation
\begin{equation*}
(I-G)\omega_{j}=I-G_{j}\,,
\end{equation*}
from which
\begin{equation*}
\omega_{ij}= (I-G_{i})\delta_{ij}+G_{i} (I-\Gamma) (I-G_{j})\,.
\end{equation*}

Note that if the operator matrix $\boldsymbol{\gamma}$
in an appropriate Banach space is bounded and its norm is less than $1$,
then
\begin{equation*}
\label{GammaSvarc}
\Gamma=\sum \Gamma_{i}-\sum_{i\ne j} \Gamma_{i}\Gamma_{j}+
\sum_{i\ne j\ne k} \Gamma_{i}\Gamma_{j}\Gamma_{k}-\ldots
\end{equation*}
where the series of sums converges in norm.
The formula precisely explains the method name.

To enclose our problem into the scheme, we will need to single out a
free resolvent
\begin{equation}
\label{svobodnaya-resolventa}
R_{0} (\lambda)= (-\Delta-\lambda I)^{-1}\,.
\end{equation}
At that
\begin{equation}
\label{redukcia}
R (\lambda)=R_{0} (\lambda) (I+\sum v_{i}R_{0} (\lambda))^{-1}\,.
\end{equation}

\section{Justification Scheme Description}
\subsection{Analytical Reasons of Existence of the Limit}
To be definite, within the framework of the ultimate absorption principle
we will follow only the passage to the limit $\mathrm{Im\,}\lambda\downarrow 0$.

As can be easily seen from~\eqref{redukcia}, we are determined to define
$G_{j} (\lambda) =-v_{j}R_{0} (\lambda)$. Assuming that
\begin{equation}
\label{R-j}
R_{j} (\lambda)= (-\Delta+v_{j}-\lambda I)^{-1}=R_{0} (\lambda) (I+v_{j}R_{0} (\lambda))^{-1}=
R_{0} (\lambda) (I-G_{j} (\lambda))^{-1}=R_{0} (\lambda) (I-\Gamma_{j} (\lambda))\,,
\end{equation}
on account of $\Gamma_{j} (\lambda) =-G_{j} (\lambda) (I-G_{j} (\lambda))^{-1}$, we find 
\begin{equation}
\label{Gamma-j-1}
\Gamma_{j} (\lambda) =v_{j}R_{j} (\lambda)\,.
\end{equation}

One should keep in mind that the Neumann series
\begin{equation*}
I-\Gamma_{j} (\lambda) =I+G_{j} (\lambda) +G_{j}^{2} (\lambda) +\ldots
\end{equation*}
are unsuitable for studying the limit $\mathrm{Im\,}\lambda\downarrow 0$,
as the powers of $G_{j} (\lambda)$ in a momentum representation accumulate a singularity
in the limit at the spectral parameter $\lambda$ is approaching the real axis.
Indeed, the operator $v_{j}$ in a momentum representation has a kernel of the form
\begin{equation*}
v_{j} ( \boldsymbol{q}, \boldsymbol{q}')= \hat v (k_{j}-k'_{j})\delta(p_{j}-p'_{j})\,,
\end{equation*}
where $\boldsymbol{q}= (k_{j},p_{j})$  is a dual to  $\boldsymbol{z}_{j}= (x_{j},y_{j})$ 
variable, $\hat v$  is in our case an
entire function and  $\delta (\cdot)$  is a delta function of Dirac. It is clear that
the kernel of the operator $G_{j}^{2} (\lambda)$ will already have singularities of the form
\begin{equation*}
\pi i\frac{\hat{v} \bigl(k_{j}-\sqrt{\lambda-p_{j}^{2}}\bigr)
\hat{v} \bigl(\sqrt{\lambda-p_{j}^{2}}-k'_{j}\bigr)
\delta (p_{j}-p'_{j})}{\sqrt{\lambda -p_{j}^{2}} \,(k'^{2}_{j}+p'^{2}_{j}-\lambda)}
\end{equation*}
(with a properly chosen branch of the square root). The situation
with a root singularity getting worse
with every new iteration. It is exactly where the main difficulty of the
mathematical scattering theory for the case of non-decreasing potentials
is concealed.

In addition, the operators $\Gamma_{j} (\lambda)$ including their continuous spectrum limit
values, can be described explicitly, as they belong to the problem
allowing a variables separation and reduction to
a one-dimensional scattering problem.

It is useful to disclose the mechanism which provides the weak limit
existence of a product $R_{j} (\lambda)= R_{0} (\lambda) (I-\Gamma_{j} (\lambda))$ at $\mathrm{Im\,}\lambda\downarrow 0$.
The free resolvent in a momentum representation on the H\"older function space
 $H^{\mu,\theta} (\mathbb{R}^{2})$ has a weak limit considered. If the function
 $v_{j}(\boldsymbol{q}, \boldsymbol{q}')$ were a smooth one \footnote{H\"olderness is sufficient} (the case of a fast
 decreasing potential in a coordinate representation), the operator  $\Gamma_{j} (\lambda)$
 would have a strong  limit and then the product $R_{0} (\lambda) (I-\Gamma_{j} (\lambda))$
 (recall that the order of multipliers for the given assertion is important)
 would have a corresponding  weak limit.

However in our case the operator $\Gamma_{j} (\lambda)$ has no strong limit mentioned.
As we will show below (section~\ref{asy-Gamma1}), the product
 $\Gamma_{j} (\lambda)\varphi\,,$ where  $\varphi\in H^{\mu,\theta}
(\mathbb{R}^{2})$ in a momentum representation,
 has a singularity of the form
\begin{equation}
\label{Gamma1-fi}
\Gamma_{j} (\lambda) \varphi(\boldsymbol{q})\sim
\frac{f (\boldsymbol{q})}{\sqrt{p_{j}^{2}-\lambda}}\,,
\end{equation}
where  $f$  is a H\"older function right up to the real value of the
parameter $\lambda$. The corresponding contribution into $(R_{j} (\lambda)\varphi,\varphi)$
takes the form 
\begin{equation}\label{suschpred}
\int dp_{j} \int dk_{j} \,\frac{f (\boldsymbol{q})
\varphi(\boldsymbol{q})}{
(p_{j}^{2}+ k_{j}^{2}-\lambda) \sqrt{p_{j}^{2}-\lambda}}
\sim  \int \frac{g (p_{j})\,dp_{j}}{p_{j}^{2}-\lambda}\,,
\end{equation}
where  $g$  is a certain H\"older function right up to the real value of
the parameter $\lambda$ (in the formula we permitted ourselves to follow only
the integral singularity). The last integral, as is known, has a limit
value at $\mathrm{Im\,}\lambda\downarrow 0$.

It is clear that in the procedure described above the analytical
character of the square root $\sqrt{p_{j}^{2}-\lambda}$ is important.

In the work we will demonstrate that also in the case of a complete
resolvent $R (\lambda)$ the ultimate absorption principle is based on a similar
analytical procedure. And exactly within the
framework of the alternating scheme one can avoid the singularity
accumulation which is seen in the Neumann series.

\subsection{Guiding Idea: Coordinate Asymptotic at Infinity of the Product $G_{i}G_{j}$ Kernel}
\label{kratkoe}
We introduce a representation of the resolvent  $R (\lambda)$
in the form of $R (\lambda)=
R_{0} (\lambda) (I-\Gamma (\lambda))$,
where  $R_{0} (\lambda)$ is a free resolvent,
while the operator  $\Gamma$ is defined by the equality 
\begin{equation*}
I-\Gamma (\lambda) = (I-G (\lambda))^{-1}\,,\quad G (\lambda)
 =\sum G_{i} (\lambda)\,,\quad G_{i} (\lambda) =v_{i}R_{0} (\lambda)\,.
\end{equation*}
The construction of the operator $\Gamma (\lambda)$ will be conducted according to the
Schwartz scheme, see above.

First of all note that the representation of the operator $G_{i}G_{j}$ in the
form of the Neumann series (when such a thing is possible), demonstrates
a usefulness of studying the square of the operator $G_{i}G_{j}$ (the iteration
method). We will apply the principle (for the reasons described above)
not to the operator $I-G$ itself, but to the matrix operator $\boldsymbol{L}$, see~\eqref{L-big}.

In our case all the operators depend on the parameter $\lambda$. We will let
ourselves omit this dependence, especially in the algebraic computations.

Denote $\boldsymbol{L}=I+\boldsymbol{\Gamma}\,$,
\begin{equation*}
  \boldsymbol\Gamma=
  \begin{pmatrix}
    0&\Gamma_{1}&\Gamma_{1}\\
\Gamma_{2}&0&\Gamma_{2}\\
\Gamma_{3}&\Gamma_{3}&0
  \end{pmatrix}\,.
\end{equation*}
Then
\begin{equation}\label{perehod-2}
\boldsymbol{L}^{-1}=(I+\boldsymbol{\Gamma})^{-1}=
(I-\boldsymbol{\Gamma}^2)^{-1}(I-\boldsymbol{\Gamma})\,.
\end{equation}
The components of operator matrix $ \boldsymbol{\Gamma}^{2}$  are certain sums of
operators of the form $\Gamma_{i}\Gamma_{j}$ at $i\ne j$.

One should expect that the key to the inversion of matrix $I-\boldsymbol{\Gamma}^{2}$ is a study of
coordinate asymptotics at infinity of the products $\Gamma_{j} \Gamma_{k} $ kernels.
A bit later we will demonstrate that an asymptotic behaviour at infinity
of the operators $\Gamma_{i}$ kernels in the case
considered bears the same character as for the operators $G_{i}$, and it in its
turn means that the character of the asymptotics mentioned for products
is easily discerned already when studying a coordinate asymptotics
at infinity of the product $G_{j} G_{k}$ kernel.

Recall that the asymptotics of the kernel
 $R_{0}(\boldsymbol{z}_{1},\boldsymbol{z}_{2}|\lambda)$ at $\mathrm{Im\,}\lambda> 0$
and  $|\boldsymbol{z}_{1}-\boldsymbol{z}_{2}|\to\infty$ is described by the relation
\begin{equation*}
  R_{0}(\boldsymbol{z}_{1},\boldsymbol{z}_{2}|\lambda+i0)\sim \frac{e^{i\pi/4}}{2\sqrt{2\pi}}
\frac{e^{i\sqrt{\lambda}|\boldsymbol{z}_{1}-\boldsymbol{z}_{2}|}}{\sqrt{|\boldsymbol{z}_{1}-\boldsymbol{z}_{2}|}}\,.
\end{equation*}
Then finding an asymptotics at infinity of the product $G_{j}(\lambda)G_{k}(\lambda)$ kernel
resolves itself to studying an asymptotics of an integral of the type
\begin{equation}\label{GG}
  v(x_{j}) \int\limits_{\mathbb{R}^{2}} \frac{v(x_{k})\,e^{i\sqrt{\lambda}
(|\boldsymbol{z_{j}}- \boldsymbol{z_{k}}|
+| \boldsymbol{z_{k}}- \boldsymbol{z}'|)}\,d \boldsymbol{z}_{k}}{\sqrt{|\boldsymbol{z_{j}}- \boldsymbol{z_{k}}|
| \boldsymbol{z_{k}}- \boldsymbol{z}'|}}
\end{equation}
at $y_{j}\to \pm\infty$. Here  $\boldsymbol{z}_{j}= (x_{j}, y_{j})$ and $\boldsymbol{z}_{k} =(x_{k},y_{k})$ are the previously introduced
Jacobi coordinates, while the variable $\boldsymbol{z}'$ (keeping in mind the character
of the alternating series) can be regarded as equal
to a Jacobi coordinate different from $\boldsymbol{z}_{k}$.

If in the integral we restrict ourselves to integrating by a sufficiently
large but finite band (i.e. by the
domain $x_{k}\in \mathrm{supp\,}v\,,\;|y_{k}|<T\,, T\gg 1$), then the asymptotics of the integral
at $y_{j}\to \pm\infty$ is easily computed on account of stabilization of the modulus $| \boldsymbol{z}_{j}-
\boldsymbol{z}_{k}|\sim |y_{j}|\pm \boldsymbol{l}_{j}\cdot \boldsymbol{z}_{k}$ (signs are correlated),
where $\boldsymbol{l}_{j}$ is a unitary vector in the axis $y_{j}$ direction. As can be easily seen,
it will 
lead to a separation of the rank two operator, the behaviour of which at $\mathrm{Im\,}\lambda\downarrow 0$
is easily controlled.
The operator left after the asymptotic separation will have a kernel
fairly fast decreasing by $\boldsymbol{z}_{j}$ at infinity, and it leads to its
compactness in the space $\hat H^{\mu',\theta'}$
at sufficiently small $\mu'>0$ and $\theta'>0$.

Now we should stress that the integral~\eqref{GG}
is to be taken only as a conditionally convergent by $y_{k}$ and this comment is
exceptionally essential for accounting a contribution into the given
integral from the domain $|y_{k}|>T$.

For the evaluation of such a contribution it is natural to resort
to integrating by parts by the variable $y_{k}$, which leads to an absolute
integral convergence and essentially reduces the matter to the separation
already considered. Indeed, if we introduce a smooth cut-off function $\chi_{T}$
equal to a unit on the interval $(-T,T)$ with a support on the interval $[-T-1,T+1]$, and put $\chi_{T}'=1-\chi_{T}$,,
then a formal integration by parts in the integral
\begin{equation}\label{GG2}
  v(x_{j}) \int\frac{\chi_{T}'(y_{k})v(x_{k})\,e^{i\sqrt{\lambda}
(|\boldsymbol{z_{j}}- \boldsymbol{z_{k}}|
+| \boldsymbol{z_{k}}- \boldsymbol{z}'|)}\,d \boldsymbol{z}_{k}}
{\sqrt{|\boldsymbol{z_{j}}- \boldsymbol{z_{k}}|
| \boldsymbol{z_{k}}- \boldsymbol{z}'|}}
\end{equation}
by the Jacobi coordinate $y_{k}$ will accumulate a power decrease either by
the variable $|\boldsymbol{z}_{j}- \boldsymbol{z}_{k}|$, or by the variable $|\boldsymbol{z}_{k}-\boldsymbol{z}'|$.

In the first case the contribution is related to asymptotic correction
terms and will be an additional term
to the compact part of the product $G_{j}G_{k}$. In the second case we can
consider (accomplishing several integrations by parts) the decrease power
by integration variable fairly large, which allows to separate a
main term of the decomposition from the multiplier $|\boldsymbol{z}_{j}-\boldsymbol{z}_{k}|^{-1/2}$
(i.e. $|\boldsymbol{z}_{j}|^{-1/2}$) at $y_{j}\to\infty$.

Though this integration by parts, at first sight, is interfered with
by the existence of a stationary point
(a geometrical optics point of reflection or transition):
\begin{equation*}
 \left( \frac{\boldsymbol{z}_{j}- \boldsymbol{z}_{k}}{|\boldsymbol{z}_{j}- \boldsymbol{z}_{k}|}
-\frac{\boldsymbol{z}_{k}- \boldsymbol{z}'}{|\boldsymbol{z}_{k}- \boldsymbol{z}'|}\right)\cdot \boldsymbol{l}_{k}=0\,,
\end{equation*}
the contribution from which must be studied separately.
However, the existence of such a stationary point
relates only to the case of a standard improper integral with which
we do not deal in the given case.
We will show below that within the framework of the alternating Schwartz
method the interpretation of the given integral as an iterated one
leads effectively to the stationary point case in a standard sense being
not realized, or, to be more exact, does not prevent a separation of
the asymptotics of the already described type.

\begin{figure}[htbp]
\label{otraj}
\begin{center}
  \includegraphics[width=12cm]{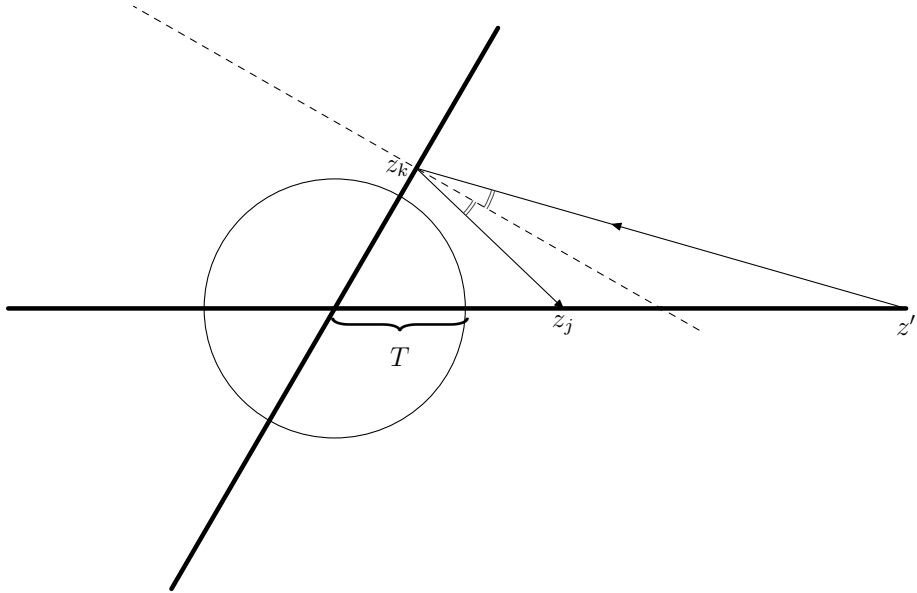}
\caption[]{To the Question of Contribution from a Stationary Point}
\end{center}
\end{figure}

To analyze the contribution from a stationary point, it is useful to
consider the off-the-screen ray reflection
geometrical optics from the standard point of view, see
picture ~\ref{otraj}. The ray transition case is analogous one.
Generally speaking, the alternating Schwartz method which we use brings
us to a study of the kernels asymptotic behaviour for chains of operators
products (in the given case of the type $G_{i_{1}}G_{i_{2}}\ldots G_{i_{k}}$) with pair-wise
differing indices of neighbouring operators. It is sufficient
for us to consider the products of threes. The matter is that as was shown
already in~\cite{BMS},\cite{BL1},
there are no more than three interactions between the screens. It means
that in the kernel of the product $G_{j}G_{k}G_{i}$ by $j\ne k\ne i$, taking into account
a stationary point in the pair $G_{j}G_{k}$, ---
define the point by $\boldsymbol{z}_{ji}$ --- the pair $G_{k}G_{i}$  can be considered as a non-having
a stationary point, which will let us integrate by parts by the variable $y_{i}$
in the corresponding kernel part of the operator $G_{j}G_{k}G_{i}$
\begin{equation}\label{triG}
   v(x_{j}) \int\limits_{\mathbb{R}^{4}} \frac{\chi(\boldsymbol{z}_{k})v(x_{k})
v(x_{i})\,e^{i\sqrt{\lambda}
(|\boldsymbol{z_{j}}- \boldsymbol{z_{k}}|+|\boldsymbol{z_{k}}- \boldsymbol{z_{i}}|
+| \boldsymbol{z_{i}}- \boldsymbol{z}'|)}\,d \boldsymbol{z}_{k}d\boldsymbol{z}_{i}}
{\sqrt{|\boldsymbol{z_{j}}- \boldsymbol{z_{k}}||\boldsymbol{z_{k}}- \boldsymbol{z_{i}}|
| \boldsymbol{z_{i}}- \boldsymbol{z}'|}}\,,
\end{equation}
where  $\chi(\boldsymbol{z})$ is a smooth finite function with a support in a fixed radius
circle and  equal identically to a unit in the vicinity of the stationary
point $\boldsymbol{z}_{ji}$.

The integration by parts leads either to an integral absolute convergence
by $\boldsymbol{z}_{k}$, which as above brings the situation to the already examined one,
i.e. to the possibility to separate an asymptotics
by $y_{j}$ of the type described above, or the integration by parts makes
the integral absolutely convergent by $\boldsymbol{z}_{i}$.
It is noteworthy that the latter again brings us to a possibility to
separate the asymptotics mentioned above,
as in the case the variable $\boldsymbol{z}_{i}$ can be regarded as defined on a compact
domain, but as results from the reflections geometry, the variable $\boldsymbol{z}_{k}$
must also be considered as defined on a compact domain,
and it allows to separate the stated asymptotic behaviour.

To show it strictly, note that considering the decrease by the variable $\boldsymbol{z}_{i}$ fairly strong
(which we achieve by iterative integration by parts), we can change the order of integration, by making the
integral by $\boldsymbol{z}_{i}$ an outer one. In the inner integral by $\boldsymbol{z}_{k}$ we can make use
of the mean value theorem version which will put the variable $\boldsymbol{z}_{k}$ in the point in the
vicinity of the reflection point $\boldsymbol{z}_{ji}$, which in its turn means that we can start decomposing
the function $|\boldsymbol{z}_{j}-\boldsymbol{z}_{k}|^{-1/2}$ by the Taylor formula  by the degrees
 $\boldsymbol{z}_{k}$, again separating the contribution into the asymptotics main term of the considered
integral.

It is clear that the argument, in fact, leads us to a necessity to study the third degree of the matrix
operator $\boldsymbol{\Gamma}$, which will be done in the justifications section but now, a bit roughly,
we restrict ourselves to the given considerations.

Thus, studying in the described key the kernel asymptotic of the operator
 $\Gamma_{i} (\lambda)\Gamma_{j} (\lambda)$ at $x\to\pm\infty$,
we obtain a separation of the operator
$\Gamma_{j} (\lambda)\Gamma_{k} (\lambda)$ into the sum
$
\Gamma_{j} (\lambda)\Gamma_{k} (\lambda) =A_{jk} (\lambda) + B_{jk} (\lambda)\,,
$
where  $A_{jk} (\lambda)$  is a rank 2 operator, containing all the ``bad'' part of the
product considered, i.e. the one which goes beyond the scope of the space $L_{2} (\mathbb{R}^{2})$
at $\mathrm{Im}\lambda\downarrow 0$, while the operator  $B_{jk} (\lambda)$ is a compact
operator in  $\hat H^{\mu',\theta'}$, strongly continuous by $\lambda$
 $\mathrm{Im}\lambda> 0$ and 
$0<c_{1}\leqslant \mathrm{Re}\lambda \leqslant c_{2}<\infty$,
if  $\mu'>0$ and $\theta'>0$ are small enough.

 The limit operator  $A_{jk} (E+i0)$ on the functions from 
 $\hat{H}^{\mu',\theta'}$ acts into a two-dimensional functions space of the form 
 \begin{equation*}
  v(x_{j})\varphi_{j}(x_{j},0)|y_{j}|^{-1/2}e^{i|y_{j}|\sqrt{E}}(C_{1}\chi_{+} (y_{j})+
C_{2} \chi_{+} (-y_{j}))
\end{equation*}
 where  $\chi_{+}(x)$  is defined as a smooth function equal identically to a unit at $x>T+1$ and
 to a zero at $x<T$. The space of such functions has root analytical singularities on the axis in the
 impulse representation, especially singularities of the form $(p\pm \sqrt{E})^{-1/2}$.
We will call the described image of the operator $A_{jk} (\lambda)$ a functions space of the type
$A_{j}$, while a vector space of such functions (by a fixed $T$) will denote by $V_{i}$.

Thus, the operator kernel $A_{jk}(\lambda)$
takes the form 
\begin{equation*}
  \Phi_{1}(\boldsymbol{z}_{1}|\lambda)
\Psi_{1}(\boldsymbol{z}_{2}|\lambda)+\Phi_{2}(\boldsymbol{z}_{1}|\lambda)
\Psi_{2}(\boldsymbol{z}_{2}|\lambda)\,,
\end{equation*}
where  $\Phi_{1},\Phi_{2}$  are functions of the type  $A_{j}$.
Note that the functions  $\Psi_{1,2}(\boldsymbol{z}_{2}|E+i0)$
at $|\boldsymbol{z}_{2}|\to\infty$
behave as diverging waves 
\begin{equation*}
\Psi_{1,2}(\boldsymbol{z}_{2}|E+i0)\sim  \frac{e^{i\sqrt{E}|\boldsymbol{z}_{2}|}}{\sqrt{|\boldsymbol{z}_{2}}|}\,.
\end{equation*}

It should be stressed that it is the necessity of a separation of such an operator as $A_{jk}$ that is an
essential distinguishing feature of the given problem --- the scattering problem of three one-dimensional
particles --- in comparison with a case of the scattering problem of three three-dimensional particles,
considered in the work ~\cite{Fad}.

The sequence of the separation $\Gamma_{i} (\lambda)\Gamma_{j} (\lambda)$  is an effective (from the alternating
scheme perspectives) representation
\begin{equation}\label{i-gamma2}
I- \boldsymbol{\Gamma}^{2} (\lambda) =I- \boldsymbol{A}(\lambda) -
 \boldsymbol{B} (\lambda)\,,
\end{equation}
where the operator  $\boldsymbol{A}$  is of a finite rank, the components of which are composed
of $A_{ij} (\lambda)$, while the components of the operator $\boldsymbol{B} (\lambda)$ are compact in
 $\hat H^{\mu',\theta'}$ with small enough positive
$\mu'$ and $\theta'$.

It is probably worth mentioning that the operators $A_{ij} (\lambda)$ or $B_{ij} (\lambda)$ are not
components of operator matrices $\boldsymbol{A} (\lambda)$ or $\boldsymbol{B} (\lambda))$, correspondingly,
in a sense of matrix analysis standard notations, but the components of the given operator matrices
are composed of the operators $A_{ij} (\lambda)$
or  $B_{ij} (\lambda)$ by the same rules, according to which the components $\boldsymbol{\Gamma}^{2}$ are
composed of the products $\Gamma_{i}\Gamma_{j}$.

\subsection{The Structure of the Operator  $I-\Gamma$}
\label{punkt-I-Gamma}

From here and from \eqref{perehod-2},\eqref{psi-gamma} by way of a complimentary reduction based on the
already described considerations, for the operator $\boldsymbol{\gamma}$ we obtain a representation
\begin{equation}
  \label{L-1}
  \boldsymbol{\gamma}=(I-\boldsymbol{\Gamma})\mathrm{diag\,}(\Gamma_{1},\Gamma_{2},\Gamma_{3})+\boldsymbol{A}_{2}+\boldsymbol{B}_{2}\,,
\end{equation}
where  $\boldsymbol{A}_{2}$ and $\boldsymbol{B}_{2}$ still possess the same properties as
$\boldsymbol{A}$ and $\boldsymbol{B}$.

As follows from ~\eqref{suschpred}, if $\varphi_{\lambda}$ is a function of the type $A_{j}$, while
$\psi\in \hat{H}^{\mu,\theta}$, then $(R_{0} (\lambda)\varphi_{\lambda},\psi)$
has a limit at $\mathrm{Im}\lambda\downarrow 0$.
As a sequence, for the desired scalar operator $\Gamma (\lambda)$ constructions of the alternating Schwartz
method lead to a representation
\begin{equation}
  \label{konechpredstGamma}
  \Gamma (\lambda) = \sum \Gamma_{i} (\lambda)-\sum_{i\ne j}\Gamma_{i}\Gamma_{j} +N (\lambda)\,,\quad
N(\lambda)=A(\lambda)+B(\lambda)\,,
\end{equation}
where  $A(\lambda)$ is a finite rank operator acting into the algebraic sum of class $A_{j}$ spaces,
while $B(\lambda)$ is a compact operator in $\hat H^{\mu,\theta}$, by $\mu,\theta>\frac{1}{2}$
(the change of these parameters range to be discussed in the strict justifications section) and has a strong
limit at $\mathrm{Im\,}\lambda\downarrow0$.
At that the operator $N (\lambda)$ possesses the property that the product $R_{0} (\lambda)N (\lambda)$ has
a weak limit in $\hat H^{\mu,\theta}$ at $\mathrm{Im}\lambda\downarrow 0$ and
$0<c_{1}\leqslant \mathrm{Re}\lambda \leqslant c_{2}<\infty$ for  $\mu,\theta>\frac{1}{2}$.
It means that
   $R (\lambda)$ has a weak limit in  $\hat H^{\mu,\theta}$ at $\mathrm{Im}\lambda\downarrow 0$ and
   $0<c_{1}\leqslant \mathrm{Re}\lambda \leqslant c_{2}<\infty$, and the limit is described by a weak equality
\begin{multline}
  \label{rezolventa-otvet}
  R(E+i0)=R_{0}(E+i0)[I-\sum \Gamma_{j}(E+i0)]\\+\sum_{i\ne j}R_{0}(E+i0)\Gamma_{i}(E+i0)\Gamma_{j}(E+i0)-\tilde{A}(E+i0)-\tilde{B}(E+i0)\\
=R_{0}(E+i0)-\sum_{i\ne j}R_{i}(E+i0)\Gamma_{j}(E+i0)-\tilde{A}(E+i0)-\tilde{B}(E+i0)\,,
\end{multline}
where  $\tilde{A}=R_{0}A$  is a finite rank operator, while
 $\tilde{B}=R_{0}B$  is ``compact''.

 From the physical point of view the operator $\tilde{A}$ describes a wave
 after at least three interactions
with screens, while $\tilde{B}$ --- a wave reflected from the center.

In the following section we will turn the ideas into strict statements.

\section{Justifications }\label{obosnov}

\subsection{Asymptotics of the Kernel $\Gamma_{j}$}
\label{asy-Gamma1}

From  considerations of symmetry it is sufficient to describe a coordinate
asymptotic of the kernel
$\Gamma_{1}(\boldsymbol{z},\boldsymbol{z}'|E+i0)$, when $\boldsymbol{z}$
lies in the first band and $\boldsymbol{z}'$ in the second.
It is easiest to regard the kernel of a "two-particles" resolvent
$R_{1}(\boldsymbol{z},\boldsymbol{z}'|\lambda)$
as a convolution of
\begin{equation*}
  R_{1}(\boldsymbol{z},\boldsymbol{z}'|\lambda)=\frac{1}{2\pi i}
\int\limits_{C}d\xi r(x,x'|\xi)r_{0}(y,y'|\lambda-\xi)
\end{equation*}
where  $r$ is a resolvent of one-dimensional problem with a potential $v(x)$,
$r_{0}$  is a corresponding free resolvent, and $C$  is a contour of
integrating around the positive semi-axis in a standard negative direction,
separating the point $\lambda$ from the positive semi-axis.
To prove the formula, introduce spectral measures $E(s)$ and $E_{0}(s)$
of the Schroedinger operators of the
corresponding one-dimensional problems, then with an application of
the residues theory
\begin{equation*}
  \frac{1}{2\pi i}\int\limits_{C}d\xi\iint\limits_{\mathbb{R}^{2}}\frac{dE(s)dE_{0}(t)}{(s-\lambda+\xi)
(t-\xi)}=\iint\limits_{\mathbb{R}^{2}}\frac{dE(s)dE_{0}(t)}{(s+t-\lambda)}=
R_{1}(\lambda)\,.
\end{equation*}

In connection with a position of the points
$\boldsymbol{z}=(x,y)$ and
$\boldsymbol{z}'=(x',y')$ we regard $x'$ and $y-y'$ as large, while
$x$ is limited.

The free resolvent $r_{0}$ for the case $\mathrm{Im\,}k>0$ is given by the equality
\begin{equation*}
r_{0}(y,y'|k^{2})=\frac{i}{2k}e^{ik|y-y'|}\,.
\end{equation*}
Next, denote by  $\varphi_{-}$  a solution of the equation
\begin{equation*}
  -\varphi''+v\varphi=k^{2}\varphi\,,
\end{equation*}
equal to  $s(k)e^{ikx}$ at large positive $x$ ($s(k)$ is transition
coefficient and even function). Analogously, $\varphi_{+}$  is a solution of
the same equation, equal to
$s(k)e^{-ikx}$ at large negative $x$. The Jacobian of the solutions is equal to
$W[\varphi_{+},\varphi_{-}]=2ik s(k)$.
The resolvent  $r (x,x'|k^{2})$ is equal to 
\begin{equation*}
  \frac{\varphi_{+}(x)\varphi_{-}(x')}{W}\,,\quad x<x'\quad\text{and}\quad
 \frac{\varphi_{-}(x)\varphi_{+}(x')}{W}\,,\quad x>x'\,.
\end{equation*}
For the sake of definiteness consider the case when
$x'$ and $y-y'$ are large positive.  Then 
\begin{multline*}
  R_{1}(\boldsymbol{z},\boldsymbol{z}'|\lambda)=\frac{-1}{2\pi i}
\int\limits_{C}dk^{2}\varphi_{+}(x,k)\frac{1}{4k\sqrt{\lambda-k^{2}}}\,
e^{ikx'+i\sqrt{\lambda-k^{2}}(y-y')}\\=
\frac{-1}{4\pi i}\int\limits_{\mathbb{R}}\frac{\varphi_{+}(x,k)\,dk}{\sqrt{\lambda-k^{2}}}\,
e^{ikx'+i\sqrt{\lambda-k^{2}}(y-y')}\,,
\end{multline*}
where the branch of the square root $\sqrt{\lambda-k^{2}}$ is chosen with
a positive imaginary component. The main contribution in the asymptotic of
the present integral is brought by the
saddle-point turning into a stationary one when the spectral parameter
$\lambda$ approaches the positive
semi-axis. The point is described by the equation
\begin{equation*}
  x'\sqrt{\lambda-k^{2}}=k(y-y')\,.
\end{equation*}
At that at the saddle-point 
\begin{equation*}
  k=\frac{\sqrt{\lambda}x'}{|\boldsymbol{z}-\boldsymbol{z}'|}+O(|\boldsymbol{z}-\boldsymbol{z}'|^{-1})\,,\quad
\sqrt{\lambda-k^{2}}=\frac{\sqrt{\lambda}(y-y')}{|\boldsymbol{z}-\boldsymbol{z}'|}+O(|\boldsymbol{z}-\boldsymbol{z}'|^{-1})\,,
\end{equation*}
and the phase function $\Phi(k)=kx'+\sqrt{\lambda-k^{2}}(y-y')$
is equal to  $\sqrt{\lambda(x'^{2}+(y-y')^{2})}=\sqrt{\lambda}
|\boldsymbol{z}-\boldsymbol{z}'|+O(1)$. The second derivative of the phase function is equal to
 \begin{equation*}
   \Phi''(k)=\frac{\lambda(y-y')}{(\lambda-k^{2})^{3/2}}
 \end{equation*}
and at the saddle-point is equal to 
\begin{equation*}
  \frac{1}{\sqrt{\lambda}}\frac{|\boldsymbol{z}-\boldsymbol{z}'|^{3}}{(y-y')^{2}}
+O(1)\,.
\end{equation*}
The saddle-point method (or the one of the stationary phase, if $\lambda$ approaches the real axis)
gives the asymptotic
\begin{equation}\label{asy-R1}
  R_{1}(\boldsymbol{z},\boldsymbol{z}'|\lambda)=
\varphi_{+}(x,k_{0})\frac{e^{i\pi/4}}{2\sqrt{2\pi}\sqrt[4]{\lambda}}\frac{e^{i\sqrt{\lambda}|\boldsymbol{z}-
\boldsymbol{z}'|}}{\sqrt{|\boldsymbol{z}-\boldsymbol{z}'|}}(1+O(|\boldsymbol{z}-
\boldsymbol{z}'|^{-1}))
\end{equation}
(with our stated assumption concerning $\lambda$).
Note that the asymptotics justifies the equation ~\eqref{Gamma1-fi},
as the asymptotics main order with an accuracy to a multiplier depending
on the variable $x$ coincides with
the free resolvent asymptotics.

 \subsection{The Asymptotics of the Kernel $\Gamma_j\Gamma_k\Gamma_{i}$}
 
 In the subsection we will essentially repeat the arguments of the
 section ~\ref{kratkoe}.

Meaning the asymptotics ~\eqref{asy-R1}, we will turn to the integral
\begin{equation}\label{triGamma}
   I=\tilde{v}(x_{j}) \int\limits_{\mathbb{R}^{4}} \frac{\tilde{v}(x_{k})
\tilde{v}(x_{i})\,e^{i\sqrt{\lambda}
(|\boldsymbol{z_{j}}- \boldsymbol{z_{k}}|+|\boldsymbol{z_{k}}- \boldsymbol{z_{i}}|
+| \boldsymbol{z_{i}}- \boldsymbol{z}'|)}\,d \boldsymbol{z}_{k}d\boldsymbol{z}_{i}}
{\sqrt{|\boldsymbol{z_{j}}- \boldsymbol{z_{k}}||\boldsymbol{z_{k}}- \boldsymbol{z_{i}}|
| \boldsymbol{z_{i}}- \boldsymbol{z}'|}}\,,
\end{equation}
compare~\eqref{triG}, where  $\tilde{v}=v\varphi_{+}$ is a finite function.
It is the integral that with an accuracy to a constant multiplier describes
the asymptotics of the kernel
 $\Gamma_{j}\Gamma_{k}\Gamma_{i}$ at $y_{j}\to\infty$.

 Thus, having fixed the large $T\gg 1$, we will make use of the separation of
 the unity
 $1=\chi_{T}(y_{k})+\chi'_{T}(y_{k})\,,\; \chi'_{T}(x)=\chi_{+}(x)+\chi_{+}(-x)$,
 described in the section~\ref{kratkoe}. In the integral
\begin{equation}\label{triGamma-1}
   I_{T}=\tilde{v}(x_{j}) \int\limits_{\mathbb{R}^{4}} \frac{\chi_{T}(y_{k})\tilde{v}(x_{k})
\tilde{v}(x_{i})\,e^{i\sqrt{\lambda}
(|\boldsymbol{z_{j}}- \boldsymbol{z_{k}}|+|\boldsymbol{z_{k}}- \boldsymbol{z_{i}}|
+| \boldsymbol{z_{i}}- \boldsymbol{z}'|)}\,d \boldsymbol{z}_{k}d\boldsymbol{z}_{i}}
{\sqrt{|\boldsymbol{z_{j}}- \boldsymbol{z_{k}}||\boldsymbol{z_{k}}- \boldsymbol{z_{i}}|
| \boldsymbol{z_{i}}- \boldsymbol{z}'|}}
\end{equation}
we easily separate the asymptotics at $y_{j}\to\infty$, as the function
$|\boldsymbol{z}_{j}-\boldsymbol{z}_{k}|^{-1/2}$ can be decomposed in terms of powers of 
$\boldsymbol{z}_{k}$. At that (omitting the dependence on $j,k,i$)
\begin{equation}
  \label{triGamma-1a}
  I_{T}=A_{T}+B_{T}\,,
\end{equation}
where  $A_{T}$  is a rank two operator with a kernel of the form 
\begin{equation}
  \label{Aranga2}
  A_{T}(\boldsymbol{z}_{j},\boldsymbol{z}')=
  v(x_{j})\varphi_{j}(x_{j},0)|y_{j}|^{-1/2}e^{i|y_{j}|\sqrt{E}}(\chi_{+} (y_{j})\psi_{+}(\boldsymbol{z}')+\chi_{+} (-y_{j})\psi_{-}(\boldsymbol{z}'))\,,
\end{equation}
the functions  $\psi_{\pm}$ satisfy the estimate
\begin{equation}
  \label{ocenca-psi-pm}
  |\psi_{\pm}(\boldsymbol{z})|\leqslant \frac{C}{\sqrt{1+|\boldsymbol{z}|}}\,,
\end{equation}
while the kernel of the operator $B_{T}$ satisfies the estimate
\begin{equation}
  \label{ocenca-BT}
  |B_{T}(\boldsymbol{z}_{j},\boldsymbol{z}')|\leqslant
\frac{C|v(x_{j})|}{(1+|\boldsymbol{z}_{j}|)^{3/2}(1+|\boldsymbol{z}'|)^{1/2}}\,.
\end{equation}
A decomposition of the form~\eqref{triGamma-1a} will be called
the decomposition of the type~$AB$.

In the integral
\begin{equation}\label{triGamma-2}
   I'_{T}=\tilde{v}(x_{j}) \int\limits_{\mathbb{R}^{4}} \frac{\chi'_{T}(y_{k})\tilde{v}(x_{k})
\tilde{v}(x_{i})\,e^{i\sqrt{\lambda}
(|\boldsymbol{z_{j}}- \boldsymbol{z_{k}}|+|\boldsymbol{z_{k}}- \boldsymbol{z_{i}}|
+| \boldsymbol{z_{i}}- \boldsymbol{z}'|)}\,d \boldsymbol{z}_{k}d\boldsymbol{z}_{i}}
{\sqrt{|\boldsymbol{z_{j}}- \boldsymbol{z_{k}}||\boldsymbol{z_{k}}- \boldsymbol{z_{i}}|
| \boldsymbol{z_{i}}- \boldsymbol{z}'|}}
\end{equation}
we will select a vicinity of the stationary point $\boldsymbol{z}_{ji}$
by the variable
$\boldsymbol{z}_{k}$, dividing the integral into a sum of integrals.
Let $I_{p}$ be the integral
in which integrating by $\boldsymbol{z}_{k}$ is carried out in this vicinity,
and $I_{o}$  be an
additional integral.

In the integral $I_{o}$ we can multiply integrate by parts by the variable $y_{k}$.
Note that at $z=x+iy$ and $z_{0}=x_{0}+iy_{0}$
\begin{align*}
  \frac{\partial}{\partial y}|z-z_{0}|^{-k/2}&=-\frac{k}{4}|z-z_{0}|^{-(k+4)/2}(y-y_{0})\,,
\\
\frac{\partial^{2}}{\partial y^{2}}|z-z_{0}|^{-k/2}&=\frac{k^{2}}{16}|z-z_{0}|^{-(k+4)/2}-
\frac{k(k+4)}{16}|z-z_{0}|^{-(k+8)/2}(x-x_{0})^{2}\,.
\end{align*}
It means that with an accuracy to a term of the type $B_{T}$ (with an estimate~\eqref{ocenca-BT})
we will come to integrals of the form
\begin{equation}\label{triGamma-2a}
   \tilde{v}(x_{j}) \int\limits_{\mathbb{R}^{4}} \frac{\chi'_{T}(y_{k})\tilde{v}(x_{k})
\tilde{v}(x_{i})\,e^{i\sqrt{\lambda}
(|\boldsymbol{z_{j}}- \boldsymbol{z_{k}}|+|\boldsymbol{z_{k}}- \boldsymbol{z_{i}}|
+| \boldsymbol{z_{i}}- \boldsymbol{z}'|)}\,d \boldsymbol{z}_{k}d\boldsymbol{z}_{i}}
{\sqrt{|\boldsymbol{z_{j}}- \boldsymbol{z_{k}}||\boldsymbol{z_{k}}- \boldsymbol{z_{i}}|^{N}
| \boldsymbol{z_{i}}- \boldsymbol{z}'|}}
\end{equation}
with fairly large positive $N$ (and a bit overridden function $\tilde{v}(x_k)$),
which allows to separate
several terms in a decomposition of the function $|\boldsymbol{z}_{j}-\boldsymbol{z}_{k}|^{-1/2}$
by the powers of $\boldsymbol{z}_{k}$. As a sequence, the integral $I_{o}$
obtains a decomposition into a sum
 of the type $AB$, with the part $B$ in the decomposition satisfying the estimate ~\eqref{ocenca-BT}
 with a constant $C$, which can be made arbitrarily small at  $T\to\infty$.

 In the integral $I_{p}$ we will change the order of integration,
 making the integral
 by $\boldsymbol{z}_{k}$ an inner one. Then in the outer integral by
 the variable $y_{i}$
 we can integrate by parts, on account of the absence for it of the
 stationary point (recall
 that for the given geometry there cannot be two reflections).
 It means that we will come to a sum
of two integrals of the form
\begin{equation}\label{triGamma-2b}
   \tilde{v}(x_{j}) \int\frac{\chi'_{T}(y_{k})\tilde{v}(x_{k})
\tilde{v}(x_{i})\,e^{i\sqrt{\lambda}
(|\boldsymbol{z_{j}}- \boldsymbol{z_{k}}|+|\boldsymbol{z_{k}}- \boldsymbol{z_{i}}|
+| \boldsymbol{z_{i}}- \boldsymbol{z}'|)}\,d \boldsymbol{z}_{k}d\boldsymbol{z}_{i}}
{\sqrt{|\boldsymbol{z_{j}}- \boldsymbol{z_{k}}||\boldsymbol{z_{k}}- \boldsymbol{z_{i}}|^{N}
| \boldsymbol{z_{i}}- \boldsymbol{z}'|^{K}}}\,,
\end{equation}
where  $N+K$ is arbitrarily large and integration by $\boldsymbol{z}_{k}$ is extended onto the
vicinity of the point $\boldsymbol{z}_{ji}$. But as it follows in an
elementary way from the geometry
of reflections, $|\boldsymbol{z}|_{ji}\leqslant C|\boldsymbol{z}_{i}|$
with a certain constant $C$, not depending on the position of the point
$\boldsymbol{z}_{j}$. It will again let us separate several decomposition terms
of the function
$|\boldsymbol{z}_{j}-\boldsymbol{z}_{k}|^{-1/2}$ into the Taylor series which proves that the integral
$I_{p}$ is also decomposed in the sum of the type $AB$.

Thus we can state that the representation of the type $AB$ is true:
\begin{equation}
  \label{prod3gamma}
\Gamma_{i}\Gamma_{j}\Gamma_{k}=A_{ijk}+B_{ijk}\,.
\end{equation}
At that the procedure of an asymptotic separation of the product considered, described above,
brings to light the following property useful for the furthest. The operator
$B_{ijk}$ can be represented in the form
\begin{equation}
  \label{eq:razlojenie-B}
  B_{ijk}=\tilde{\Gamma}_{i}\tilde{\Gamma}_{j}\tilde{\Gamma}_{k}+E_{ijk}\,,
\end{equation}
where  $\tilde{\Gamma}_{i}$ has the same sense as the operator $\Gamma_{i}$
but for
a cut-off potential $v_{i}\chi_{T}$, i.e. for a finite one, and $E_{ijk}$ satisfy
the estimates
~\eqref{ocenca-BT}, in which the constant $C$ can be made arbitrarily small at $T\to\infty$.

\subsection{Inversion of $I+\boldsymbol{\Gamma}^{3}$}

Meaning the equality
\begin{equation}
  \label{Gamma1-3}
  (I+\boldsymbol{\Gamma})^{-1}=(I-\boldsymbol{\Gamma}+\boldsymbol{\Gamma}^{2})
(I+\boldsymbol{\Gamma}^{3})^{-1}\,,
\end{equation}
note that the procedure of the construction of the operator $I-\Gamma$ is based now on the inversion
of the matrix operator $I+\boldsymbol{\Gamma}^{3}$ with a control of the transfer to the limit at
$\mathrm{Im\,}\lambda\downarrow 0$ on the appropriate functions space.

From the equation ~\eqref{prod3gamma} there follows a representation
\begin{equation}\label{i-gamma3}
I+ \boldsymbol{\Gamma}^{3} (\lambda) =I- \boldsymbol{A}(\lambda) -
 \boldsymbol{B} (\lambda)\,,
\end{equation}
where the operator  $\boldsymbol{A}$  is of the finite rank, the components of which
are combined from $A_{ijk} (\lambda)$, while the components of the operator $\boldsymbol{B} (\lambda)$
 are from $B_{ijk}$. The estimate~\eqref{ocenca-BT},  true for the components of the operator
$\boldsymbol{B}$, demonstrate that the operator acts into the space of the functions 
$\hat H^{\mu,\theta}$
at $\mu<\frac{1}{2}$. To be able to regard the operator as a compact one, we must
proceed from
the space of the functions $\hat H^{\mu,\theta}$ with sufficiently small $\mu$ and
$\theta$.
Unfortunately, the operators $\Gamma_{i}$ in such spaces are not defined, to be more exact, we should
require $\mu,\theta>\frac{1}{2}$. We will be helped by the fact that the operator acts into the space
of the functions with a support in a band, and to search the inverse to $I+\boldsymbol{\Gamma}^{3}$
is sufficient exactly for such functions.

Indeed, for $f\in L_{q}(\mathbb{R}^{2})$ on account of the H\"older inequality
\begin{equation*}
  \Bigl|\int\limits_{\mathbb{R}^{2}}(1+|\boldsymbol{z}|)^{-1/2}f(\boldsymbol{z})\,d\boldsymbol{z}
\Bigr|\leqslant \left(\int\limits_{\mathbb{R}^{2}}(1+|\boldsymbol{z}|)^{-q'/2}d\boldsymbol{z}\right)^{1/q'}
\left(\int\limits_{\mathbb{R}^{2}}|f(\boldsymbol{z})|^{q}d\boldsymbol{z}\right)^{1/q}<\infty
\end{equation*}
at $q'>4$ ($q$ and $q'$  are conjugate exponents), i.e. it is necessary that
$q<\frac{4}{3}$. From the theorem ~\ref{thmLqH} (the second statement of the theorem) it follows that
$f$ from $\hat H^{\mu,\theta}$ lies in $L_{q}$ for a certain $q<\frac{4}{3}$, if $\mu,\theta>\frac{1}{2}$.
It is exactly the space that we fix.

The operator $\Gamma_{i}$ at the limit value of the parameter $\lambda$ maps
the function from
$\hat H^{\mu,\theta}$ in the function from $\hat H^{\nu,\theta}$ for $\nu>0$,
but at that the support of
the latter lies in the band defined by the finite space of the potential $v$.
In the Fourier images
the action of the operator $\Gamma_{i}$ on the function $\hat\varphi\in H^{\mu,\theta}$ will be defined
by the equality
\begin{multline}\label{gamma-i-fi}
  \Gamma_{i}\hat\varphi(k,p)=\iint dk'dp'\delta(p-p')\tilde{v}(k-k')\frac{\hat\varphi(k',p')}{p'^{2}
+k'^{2}-\lambda}\sim \frac{f(k,p)}{\sqrt{p^{2}-\lambda}}\\=\frac{f(k,\sqrt{\lambda})\eta(p)}{
\sqrt{p^{2}-\lambda}}+\frac{f(k,p)-f(k,\sqrt{\lambda})\eta(p)}{\sqrt{p^{2}-\lambda}}\,,
\end{multline}
where $f(p,k)=\pi i\tilde{v}(k-\sqrt{p^{2}-\lambda})\hat\varphi(\sqrt{p^{2}-
\lambda},p)$, $k,p$  are dual variables relative to $x_{i},y_{i}$,
the function  $\tilde{v}$ is (with an accuracy to a constant multiplier)
a Fourier image of the product
$v(x)\varphi_{+}(x,k_{0})$, the function $\eta$ is a smooth finite one,
identically equal to a unit in
the vicinity of the point $\sqrt{\lambda}$.

From the representation there follows that $\Gamma_{i}\hat H^{\mu,\theta}$
is a direct sum of spaces,
one of which is two-dimensional and consists of the functions of $A_{i}$ class,
while the second one contains functions from
$\hat H^{\nu,\theta}\,,\;\nu=\mu-\frac{1}{2}>0\,,$ with a support in a band.
We will denote the latter by
 $\hat H^{\nu,\theta}_{i}$, and then $\Gamma_{i}\hat H^{\mu,\theta}\subset
V_{i}+\hat H^{\nu,\theta}_{i}$. At the limit values of the parameter $\lambda$ 
the sum is direct.

From the first part of theorem ~\ref{thmLqH} there follows that the function 
from $\hat H^{\mu,\theta}_{i}$
lies in $L_{r}$ for $r<2$  in the direction of the $i$-band. 
Thus the operator $\boldsymbol{\Gamma}^{3}$
will be regarded as an operator in the space $\boldsymbol{W}^{\nu,\theta}=\oplus_{i} (V_{i}+\hat H^{\nu,\theta}_{i})$.

To inverse the operator $I+\boldsymbol{\Gamma}^{3}$ we will use again the 
alternating Schwartz procedure.

First, the inversion of the operator $I-\boldsymbol{A} (\lambda)$ is
 controlled explicitly, as
$\boldsymbol{A}(\lambda)$  is a finite rank operator. As we remember, 
the construction of the operator
depended on the large parameter $T$. It is easy to verify that at 
sufficiently large $T$
inverse operator $(I- \boldsymbol{A} (\lambda))^{-1}=I-\boldsymbol{\Gamma}_{A} (\lambda)$ exists and
has a formal finite operator at $\mathrm{Im\,}\lambda\downarrow 0$, acting into an algebraic sum
of spaces of the type  $A_{j}\,,\,j=1,2,3$, i.e. into the space
$\boldsymbol{V}=\oplus_{i}V_{i}$. It is proved in the appendix~\ref{A}.

Second, as was already noted, the operator $\boldsymbol{B} (\lambda)$ has the form
 $\boldsymbol{B} (\lambda)=\tilde{\boldsymbol{\Gamma}}^{3} (\lambda) +
\boldsymbol{E} (\lambda)$, where the operator $\tilde{\boldsymbol{\Gamma}} (\lambda)$ 
has exactly the same sense
as the operator $\boldsymbol{\Gamma} (\lambda)$, but for the finite potential
$\tilde{V}=V\cdot \chi_{T}$, where $\chi_{T}$  is an arbitrary smooth finite function equal to a unit in the
central circle of the radius $T$ and having a support in the central circle
of radius $T$,
while the operator $\boldsymbol{E} (\lambda)$  is a small norm operator, if $T$ is large enough.

The operator $\boldsymbol{B}$ acts from
$\boldsymbol{W}^{\nu,\theta}$ in $\oplus_{i}\hat H^{\sigma,\tau}_{i}$,
where $\sigma<\frac{1}{2}\,,\;\tau>\frac{1}{2}\,.$
At $\nu<\sigma\,,\;\theta<\tau$ it is compact.

Note now that the operator $\tilde{\boldsymbol{\Gamma}}^{3}$ does not have $-1$ as its eigenvalue right up to
the limit values $\lambda$ on the positive semi-axis. Moreover, a certain
vicinity $|w+1|<\delta$ of the point
 $-1$ will not contain the eigenvalues of the operator $\tilde{\boldsymbol{\Gamma}}^{3}$.
 Indeed, first note that on account of the equality 
\begin{equation*}
  (H_{0}-\lambda)(H_{0}+\tilde{V}-\lambda-i0)^{-1}=I-\tilde{\Gamma}(\lambda+i0)=(I-\tilde{G}(\lambda+i0))^{-1}
\end{equation*}
the compact operator  $\tilde{G}$ does not have a unit as its eigenvalue. 
Moreover, on account of
\begin{equation*}
  wI-\tilde{G} (\lambda) =wI-\tilde{V}R_{0}(\lambda)=
w(I-w^{-1}\tilde{V}R_{0} (\lambda))\,,
\end{equation*}
the operator $\tilde{G}(\lambda+i0)$ does not have eigenvalues in a certain vicinity of the unit,
as the only point of its spectrum is a zero. As a sequence, the operator $\tilde{\Gamma} (\lambda+i0)$,
analytically dependent on $\tilde{G} (\lambda+i0)$, does not have a unit as its eigenvalue with
a whole 
vicinity and, indeed, the only point of its spectrum is zero.

On account of the alternating scheme the operator $\tilde{\boldsymbol{\Gamma}}$ has a zero point as the 
only point of spectrum. Indeed, if the point $-1$ were the point of spectrum for the operator, then for 
certain functions $\phi_{1},\phi_{2},\phi_{3}$ from $L_{2}$ we would have 
\begin{equation*}
  \tilde{\Gamma}_{i}(\phi_{j}+\phi_{k})=-\phi_{i}\,,
\end{equation*}
where the indices $i,j,k$ form all even permutations. From here 
\begin{equation*}
  \tilde{G_{i}}\sum\phi_{j}=\phi_{i}\quad\text{and}\quad \tilde{G}\phi=\phi\,,
\end{equation*}
where $\phi=\sum\phi_{j}$, which cannot be. If $w\ne 0$ were the eigenvalue of 
$\tilde{\boldsymbol{\Gamma}}$,
then the change $\tilde{\Gamma}_{i}$ for $-w^{-1}\tilde{\Gamma}_{i}$ would reduce it to the previous one
(would bring us to the previous case).  

But then the operator $\tilde{\boldsymbol{\Gamma}}^{3}$ also has as its eigenvalue only the point zero,
which exactly proves the statement made above.   

As the operator norm $\boldsymbol{E}$ can be made arbitrarily small at large $T$, from here
there arises 
an invertibility $I-\boldsymbol{B}$. Thus, $I-\boldsymbol{B} (\lambda)$ has an inverse 
$I-\boldsymbol{\Gamma}_{B} (\lambda)$.
Note that the inverse one has a limit in a strong operator sense at $\mathrm{Im\,}\lambda\downarrow 0\,.$ 

Finally, the product 
$\boldsymbol{\Gamma}_{A} (\lambda)
\boldsymbol{\Gamma}_{B} (\lambda)$ is once again the finite rank operator and the invertibility of the operator
$I-\boldsymbol{\Gamma}_{A} (\lambda)
\boldsymbol{\Gamma}_{B} (\lambda)$ does not require an involvement of new arguments in comparison with
the invertibility of the operator $I-\boldsymbol{A} (\lambda)$, i.e. is a sequence of the considerations
in appendix ~\ref{A}.

According to the Schwartz method, we can conclude that at
  $\mathrm{Im}\lambda>0$ the operator  $(I+\boldsymbol{\Gamma}^{3})^{-1}$ has a representation
  \begin{equation}
  \label{predstGamma2Obrat}
  (I+ \boldsymbol{\Gamma}^{3})^{-1}=I-\boldsymbol{A}_{1} (\lambda)
-\boldsymbol{B}_{1} (\lambda)\,,
\end{equation}
that is a decomposition of the type $AB$, i.e. the operators 
 $\boldsymbol{A}_{1}$ and $\boldsymbol{B}_{1}$ have the same properties as the operators 
$\boldsymbol{A}$ and $\boldsymbol{B}$, the matrix components 
$\boldsymbol{A}_{1} (\lambda)$ are the finite rank operators, acting into the algebraic sum of spaces
 of the type $A_{j}\,,j=1,2,3\,,$ , while the matrix components $\boldsymbol{B}_{1}$ are compact operators
 in $\boldsymbol{W}^{\nu,\theta}$, strongly continuous by $\lambda$ at
 $\mathrm{Im}\lambda\geqslant 0$ and $0<c_{1}\leqslant \mathrm{Re}\lambda \leqslant c_{2}<\infty$,
if $\nu$ and $\theta$ are small enough.

\subsection{The Structure of the Operators $I-\Gamma$ and $R(\lambda)$}
Now we can proceed to a clarification of the provisions of the
section~\ref{punkt-I-Gamma}. 
From the correlations~\eqref{Gamma1-3},~\eqref{i-gamma3} we easily deduce that 
\begin{equation*}
  \boldsymbol{\gamma}=(I-\boldsymbol{\Gamma})\mathrm{diag}(\Gamma_{1},\Gamma_{2},\Gamma_{3})
-\boldsymbol{N}\,,
\end{equation*}
where $\boldsymbol{N}$ allows a decomposition of the $AB$ type. It fully justifies
the representation~\eqref{L-1} and, then, all the subsequent statements of the
section ~\ref{punkt-I-Gamma}. 

It should be noted that the known effect of the wave function scattering amplitude discontinuity after 
re-propagations off the screens, see ~\cite{BMS}, is an effect of the second
rank --- it is completely 
controlled by the term $\sum_{i\ne j}R_{i}\Gamma_{j}$ in the formula~\eqref{rezolventa-otvet}.

\appendix
\section{Inversion of the Finite-Dimensional Operator }\label{A}

We will consider here a procedure of a transformation of the operator 
$$
A=
\begin{pmatrix}
I-\varphi_1\langle\psi_{12},*\rangle-\varphi_1\langle\psi_{13},*\rangle &
      -\varphi_1\langle\psi_{13},*\rangle & -\varphi_1\langle\psi_{12},*\rangle\\
-\varphi_2\langle\psi_{23},*\rangle &
I-\varphi_2\langle\psi_{21},*\rangle-\varphi_2\langle\psi_{23},*\rangle &
  -\varphi_2\langle\psi_{21},*\rangle \\
-\varphi_3\langle\psi_{32},*\rangle & -\varphi_3\langle\psi_{31},*\rangle &
  I-\varphi_3\langle\psi_{31},*\rangle-\varphi_3\langle\psi_{32},*\rangle
  \end{pmatrix}
  $$
By  $A_1$ we will denote $I-A$ and instead of the operator $A^{-1}$
we will search the operator 
 $W=A^{-1}-I$, defined by the equation
$$
A_1W=W-A_1.
$$
At that the operator $A_1$ has the form of 
$$
A_1=
\begin{pmatrix}
\varphi_1\langle\psi_{12},*\rangle+\varphi_1\langle\psi_{13},*\rangle &
      \varphi_1\langle\psi_{13},*\rangle & \varphi_1\langle\psi_{12},*\rangle\\
\varphi_2\langle\psi_{23},*\rangle &
\varphi_2\langle\psi_{21},*\rangle+\varphi_2\langle\psi_{23},*\rangle &
  \varphi_2\langle\psi_{21},*\rangle \\
\varphi_3\langle\psi_{32},*\rangle & +\varphi_3\langle\psi_{31},*\rangle &
  \varphi_3\langle\psi_{31},*\rangle+\varphi_3\langle\psi_{32},*\rangle
  \end{pmatrix}
$$
It is natural to search the operator $W$ in the form of:
 $$
W=
\begin{pmatrix}
\varphi_1\alpha_{11} &  \varphi_1\alpha_{12} & \varphi_1\alpha_{13} \\
\varphi_2\alpha_{21} &  \varphi_2\alpha_{22} & \varphi_2\alpha_{23} \\
\varphi_3\alpha_{31} &  \varphi_3\alpha_{32} & \varphi_3\alpha_{33}
 \end{pmatrix}
$$
The functionals  $\alpha_{ij}\, (i,j=1,2,3)$ are to be determined. 
To be definite, we will write a system of equations for the first column elements.
\begin{align*}
  \langle\psi_{12},\varphi_1\rangle\alpha_{11}+
\langle\psi_{13},\varphi_1\rangle\alpha_{11}+
\langle\psi_{13},\varphi_2\rangle\alpha_{21}+
\langle\psi_{12},\varphi_3\rangle\alpha_{31}&=
\alpha_{11}-\langle\psi_{12},*\rangle-\langle\psi_{13},*\rangle,\\
\langle\psi_{23},\varphi_1\rangle\alpha_{11}+
\langle\psi_{21},\varphi_2\rangle\alpha_{21}+
\langle\psi_{23},\varphi_2\rangle\alpha_{21}+
\langle\psi_{21},\varphi_3\rangle\alpha_{31}&=
\alpha_{21}-\langle\psi_{23},*\rangle,\\
\langle\psi_{32},\varphi_1\rangle\alpha_{11}+
\langle\psi_{31},\varphi_2\rangle\alpha_{21}+
\langle\psi_{31},\varphi_3\rangle\alpha_{31}+
\langle\psi_{32},\varphi_3\rangle\alpha_{31}&=
\alpha_{31}-\langle\psi_{32},*\rangle.
\end{align*}

Analogous are the systems of equations for functionals $\alpha_{j2}\,(j=1,2,3)$ and
$\alpha_{j3}\,(j=1,2,3)$. For our appendices these systems are solvable on account of 
smallness of the coefficients --- scalar products --- at large $T$.

\section{$L_{p}$ The Estimation of the H\"older Functions Fourier Images}
\begin{thm}\label{thmLqH}
Let 
\begin{equation}
\label{LpH}
\int\limits_{\mathbb{R}^{n}}|f (x+h)-f (x)|^{p}\,dx=O (|h|^{\mu p})\,,\qquad
1<p<2\,,
\quad \frac{1}{p}-\frac{1}{2}<\mu<1\,,
\end{equation}
then  $\hat f (\xi)$ lies in $L_{q}$ in any direction by 
$\frac{p}{p+p\mu-1}\leqslant q<2\,.$
If~\eqref{LpH} is satisfied at
\begin{equation}
\label{LpH2}
\min \Bigl\{1, \frac{2n}{2+n}\Bigr\}<p<2\,,
\quad n  \left( \frac{1}{p}-\frac{1}{2} \right)<\mu<1\,,\quad n\geqslant 2,
\end{equation}
then  $\hat f\in L_{q} ( \mathbb{R}^{n})$ at $\frac{pn}{pn+\mu p-n}\leqslant q<2\,.$
\end{thm}

Indeed, the Fourier transform (by $x$) of the function $f (x+h)$ equals to
$e^{i\xi\cdot h}\hat f (\xi)$ and from the boundness of the Fourier operator
$L_{p}\to L_{p'}$ ($p'$  is a conjugate exponent) 
we deduce 
\begin{equation*}
\int\limits_{\mathbb{R}^{n}}|\sin \tfrac{\xi\cdot h}{2}|^{p'}|\hat f (\xi)|^{p'}\,d\xi
=O (|h|^{\mu p'})\,.
\end{equation*}
Note that $|t|\leqslant C|\sin t|$ at $|t|< \frac{\pi}{2}$, from here
\begin{equation*}
\int\limits_{|\xi\cdot h|<1}|\xi \cdot h|^{p'}|\hat f (\xi)|^{p'}\,d\xi
=O (|h|^{\mu p'})\,.
\end{equation*}
As a sequence
\begin{equation*}
\int\limits_{0}^{1/h_{j}}d\xi_{j}\, \xi_{j}^{p'} \int\limits_{\mathbb{R}^{n-1}}d\tilde\xi_{j}
|\hat f(\xi)|^{p'}=O (h_{j}^{(\mu-1)p'})\,,
\end{equation*}
where  $\tilde\xi_{j}$  is supplementary to $\xi_{j}$ component of the argument
 $\xi$
and, for definiteness,  $h_{j}>0\,.$ Symmetrizing the estimate, we will write 
\begin{equation*}
\int\limits_{0}^{\xi_{1}}d\eta_{1}\ldots \int\limits_{0}^{\xi_{n}}\,d\eta_{n}
|\eta|^{p'}|\hat f(\eta)|^{p'}=O(|\xi|^{(1-\mu)p'})
\end{equation*}
(counting for definiteness all $\xi_{j}>0$).
Determine an auxiliary function 
\begin{equation*}
\varphi (\xi)=\int\limits_{1}^{\xi_{1}}d\eta_{1}\ldots \int\limits_{1}^{\xi_{n}}\,d\eta_{n}
|\eta|^{q}|\hat f(\eta)|^{q}\,.
\end{equation*}
At that according to the H\"older inequality
\begin{equation*}
\varphi (\xi) \leqslant\left(
\int\limits_{1}^{\xi_{1}}d\eta_{1}\ldots \int\limits_{1}^{\xi_{n}}\,d\eta_{n}
|\eta|^{p'}|\hat f(\eta)|^{p'}\right)^{q/p'}\prod \xi_{j}^{1- q/p'}\leqslant
C |\xi|^{(1-\mu)q}\prod \xi_{j}^{1- q/p'}\,.
\end{equation*}

Then
\begin{multline}
\label{LpH3}
\int\limits_{1}^{\xi_{1}}d\eta_{1}\ldots \int\limits_{1}^{\xi_{n}}\,d\eta_{n}
|\hat f (\eta)|^{p'}=
\int\limits_{1}^{\xi_{1}}d\eta_{1}\ldots \int\limits_{1}^{\xi_{n}}\,d\eta_{n}
\frac{1}{|\eta|^{q}} \frac{\partial^{n}\varphi (\eta)}
{\partial \eta_{1}\ldots \partial\eta_{n}}\\= \frac{\varphi (\xi)}{|\xi|^{q}}+
\sum (\pm)\int\limits_{1}^{\xi_{j_{1}}}d\eta_{j_{1}}\ldots
\int\limits_{1}^{\xi_{j_{k}}}\,d\eta_{j_{k}}
\varphi (\zeta)
\frac{\partial^{k}}
{\partial \eta_{j_{1}}\ldots \partial\eta_{j_{k}}}\frac{1}{|\zeta|^{q}}\,,
\end{multline}
where a summation is carried out by all samples
$(j_{1},\ldots j_{k})\,,\;1\leqslant k\leqslant n\,$, at that
$\zeta_{i}=\eta_{i}$ at $i$ is equal to one of  $j_{l}$ in the considered sample and 
to $\zeta_{i}=\xi_{i}$ in the contrary case. Each term in the right-hand part~\eqref{LpH3} has the order
\begin{equation*}
O(|\xi|^{-\mu q}\prod\xi_{j}^{1-q/p'})\,.
\end{equation*}

At $1-\frac{q}{p'}-\mu q\leqslant 0 $, i.e. at $q\geqslant \frac{p}{p+p\mu-1}\,,$
the integral in the left-hand part~\eqref{LpH3} by any one separated variable $\eta_{j}$ 
remains bounded.   

At $n (1-\frac{q}{p'})-\mu q\leqslant0$,
i.e. at
 $q\geqslant \frac{pn}{pn+\mu p-n}$, 
the whole even integral in the left in~\eqref{LpH3} remains bounded.
The cases of other signs of variables $\xi_{j}$ are analogous. The conditions on $p$ and $\mu$ 
in~\eqref{LpH} and \eqref{LpH2} provide a non-emptiness of the variation
intervals $q$. The theorem is proved.







\end{document}